\documentclass[twocolumn,preprintnumbers,prb,amsmath,amssymb]{revtex4-1}

\usepackage{graphicx}
\usepackage{dcolumn}
\usepackage{bm}
\usepackage{bbm}

\usepackage{hyperref}

\begin{document}
\title{Quantitative stray field imaging of a magnetic vortex core}
\author{J.-P.~Tetienne$^{1,2}$, T. Hingant$^{1,2}$, L. Rondin$^{2}$, S.~Rohart$^3$, A.~Thiaville$^3$, J.-F.~Roch$^1$ and V.~Jacques$^{1,2}$}
\email{vjacques@ens-cachan.fr}
\affiliation{$^{1}$Laboratoire Aim\'e Cotton, CNRS, Universit\'e Paris-Sud and ENS Cachan, 91405 Orsay, France \\
$^{2}$Laboratoire de Photonique Quantique et Mol\'eculaire, Ecole Normale Sup\'erieure de Cachan and CNRS UMR 8537, 94235 Cachan, France \\
$^{3}$Laboratoire de Physique des Solides, Universit\'e Paris-Sud and CNRS UMR 8502, 91405 Orsay, France}

\begin{abstract}

Thin-film ferromagnetic disks present a vortex spin structure whose dynamics, added to the small size ($\approx10$~nm) of their core, earned them intensive study.  Here we use a scanning nitrogen-vacancy (NV) center microscope to quantitatively map the stray magnetic field above a $1 \ \mu$m-diameter disk of permalloy, unambiguously revealing the vortex core.  Analysis of both probe-to-sample distance and tip motion effects through stroboscopic measurements, allows us to compare directly our quantitative images to micromagnetic simulations of an ideal structure. Slight perturbations with respect to the perfect vortex structure are clearly detected either due to an applied in-plane magnetic field or imperfections of the magnetic structures. This work demonstrates the potential of scanning NV microscopy to map tiny stray field variations from nanostructures, providing a nanoscale, non-perturbative detection of their magnetic texture.

\end{abstract}

\maketitle

The magnetic vortex structure is an equilibrium magnetization configuration frequently found in patterned nanostructures made of soft ferromagnetic materials. It is characterized by a curling in-plane magnetization and a vortex core where the magnetization points out of the plane~\cite{Hubert}. The small size of the vortex core, on the order of $10$ nm, has been the main motivation to understand its dynamics~\cite{Choe2004,Antos2008} and harness it for applications, {\it e.g.} in the context of nanoscale magnetic memory devices and microwave generation~\cite{Drews2009,Pigeau2010,Pribiag2007}. However, it also makes it difficult to measure or image directly. It was only in 2000 that Shinjo {\it et al.} first imaged the vortex core in microdisks of permalloy using magnetic force microscopy~\cite{Shinjo2000}. Since then, several imaging techniques have enabled real-space observation of magnetic vortex cores~\cite{Wachowiak2002,Fischer2012}. However, a direct, quantitative measurement of the stray field above a vortex core has never been reported so far due to the lack of a quantitative imaging technique with sufficient spatial resolution.\\
\indent In this work we use a single nitrogen-vacancy (NV) center in diamond coupled to an atomic force microscope (AFM), {\it i.e.} a scanning NV-center microscope~\cite{Maze2008,Balasubramanian2008,Rondin2012,Maletinsky2012}, to quantitatively map the stray magnetic field above the core of a vortex state in a microdisk of permalloy. Direct comparisons with micromagnetic simulations reveal small discrepancies from an ideal vortex structure, which are attributed to imperfections of the magnetic structures. Applying a weak in-plane bias magnetic field, we then move the vortex core by few tens of nm and demonstrate how such a tiny deviation can be easily detected through modification of the stray field distribution. Beyond reporting the first quantitative maps of the stray field distribution above a magnetic vortex core, this work demonstrates how scanning-NV magnetometry can be used to detect weak perturbations of magnetic textures.     \\ 
\begin{figure*}[t]
\begin{center}
\includegraphics[width=0.92\textwidth]{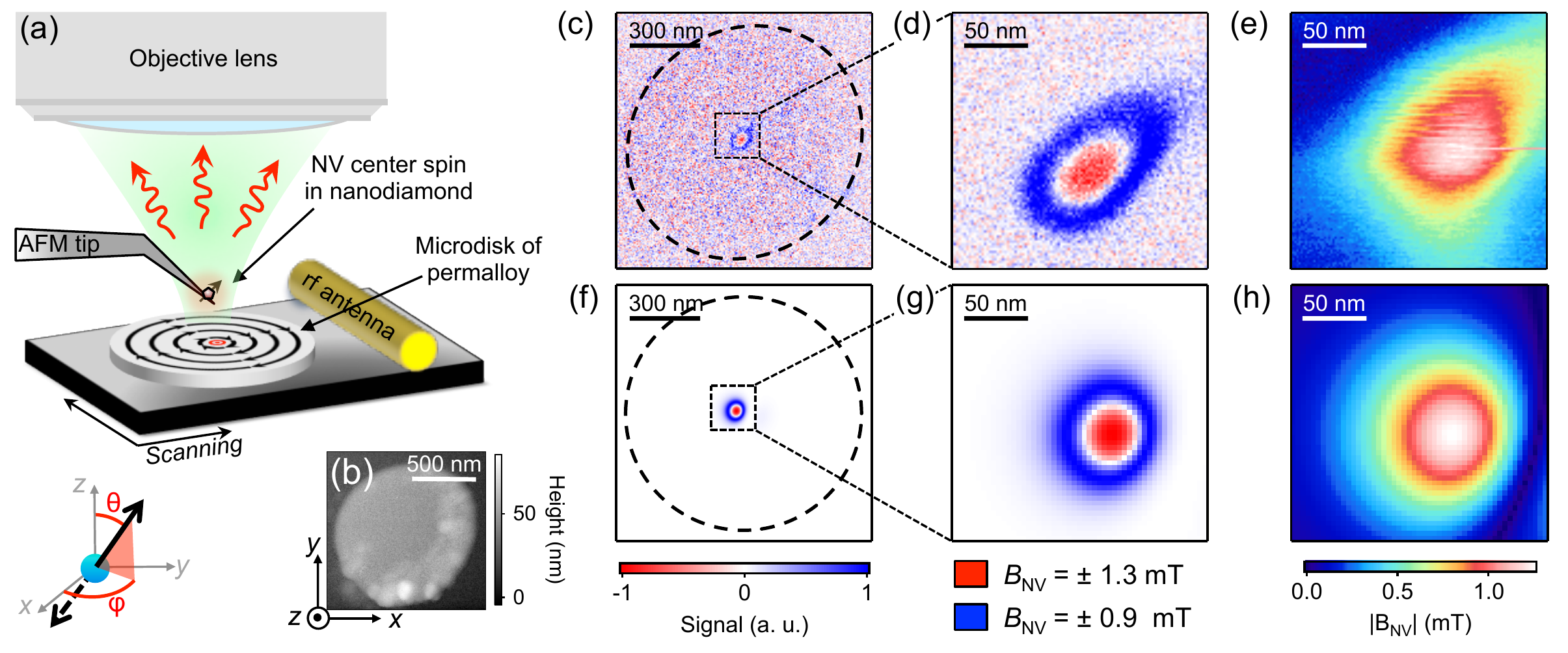}
\caption{(color online) (a) Schematic of the scanning NV-center microscope which combines an atomic force microscope (AFM) and a confocal microscope. The AFM tip is functionalized with a 40-nm nanodiamond hosting a single NV center and a radiofrequency (RF) antenna is used to perform optical detection of the NV defect ESR transition. The NV sensor used in this work has spherical angles $\theta=59^{\circ}$ and $\varphi=-15^{\circ}$ in the $xyz$ reference frame. (b) AFM image of the magnetic sample which consists of 1-$\mu$m-diameter, 50-nm-thick disks of permalloy. (c) Dual-iso-B image recorded with the scanning NV magnetometer above the disk of permalloy. The dashed circle depicts the ferromagnetic dot boundary, as extracted from the AFM image. Positive signal (blue) indicates resonance at $2895$~MHz, corresponding to $B_{\rm NV}=\pm 0.9$ mT, while negative signal (red) indicates resonance at 2905 MHz, corresponding to $B_{\rm NV}=\pm 1.3$ mT. (d) High-resolution image of the dot center. (e) Full magnetic field distribution above the dot center. (f) Simulated dual-iso-B image for a perfect disk under the same conditions as in (c). (g) Magnified view of the dot center. (h) Simulated magnetic field map for a perfect disk under the same conditions as in (e). Integration time per pixel: 60 ms in (c), 300 ms in (d), 100 ms in (e).} 
\label{Fig1}
\end{center}
\end{figure*}

A scheme of the experiment is shown in Figure~\ref{Fig1}(a). The scanning NV-center microscope makes use of an AFM operating in tapping mode, whose tip is functionalized with a $40$-nm diamond nanocrystal hosting a single NV center. This defect, which has a paramagnetic ground state ($S=1$) with a zero-field splitting $D\approx 2.87$~GHz, can be considered as a single electronic spin nestled in the diamond matrix. Importantly, it exhibits a spin-dependent photoluminescence (PL) under optical illumination, which enables measurements of the electron spin resonance (ESR) by optical means~\cite{Gruber_Science1997}. The electron spin of the NV center then serves as an atomic-size magnetic sensor which provides quantitative magnetic field imaging by recording Zeeman shifts of the NV defect electron spin sublevels~\cite{Maze2008,Balasubramanian2008,Rondin2012,Maletinsky2012}. More precisely, the ESR frequency is given by $\nu_r=D \pm g\mu_B |B_{\rm NV}|$, where $g\mu_B\approx 28$ GHz/T and $|B_{\rm NV}|$ is the magnetic field projection along the NV defect quantization axis. Experimentally, the ESR frequency is probed  with a radiofrequency (RF) field while collecting the spin-dependent PL with a confocal microscope placed on top of the AFM tip~\cite{Rondin2012}. All the experiments presented in this article were performed with the same NV sensor, whose quantization axis forms spherical angles $\theta=59^{\circ}$ and $\varphi=-15^{\circ}$ in the sample reference frame [Fig. \ref{Fig1}(b)]. These angles were inferred by recording the ESR frequency as a function of the amplitude and orientation of a calibrated magnetic field. The magnetic sample, which is composed of thermally evaporated disks of permalloy (Ni$_{80}$Fe$_{20}$), with a thickness $t=50$~nm and a $1$-$\mu$m diameter, were prepared by electron-beam lithography and lift-off on a silicon substrate. The equilibrium magnetization of these ferromagnetic microstructures is a vortex state. \\

\indent The stray magnetic field above the vortex structure was first studied with the scanning NV microscope operating in `dual-iso-B' imaging mode. This method provides magnetic images exhibiting iso-magnetic-field contours corresponding to two different values of $|B_{\rm NV}|$~\cite{Rondin2012,Rondin2013}. For instance, Figure~\ref{Fig1}(c) shows the `dual-iso-B' image recorded simultaneously with the AFM image shown in Figure~\ref{Fig1}(b), with iso-B contours corresponding to $0.9$~mT and $1.3$~mT. Since the flux-closed vortex structure does not radiate any stray field, the only pattern observed in the magnetic image is due to the vortex core at the center of the disk, where the magnetization points out of the plane. This is in contrast with a previous work on ferromagnetic square dots, where the stray field produced by N{\'e}el walls at the square diagonals overwhelms that of the vortex core~\cite{Rondin2013}. Figure~\ref{Fig1}(d) shows a high-resolution `dual-iso-B' image of the disk center, which unambiguously reveals the vortex core. In order to obtain the full magnetic field distribution above the vortex core, the value of $|B_{\rm NV}|$ was then measured at each pixel of the scan by employing a lock-in method which enables real-time tracking of the ESR frequency, as explained in details in Refs.~\cite{Schoenfeld2011,Rondin2012}. This method provides a fully quantitative map of the stray field produced by the vortex core, as illustrated in Fig. \ref{Fig1}(e). We note that the stray field spreads over $\approx 100$~nm, which corresponds to the resolving power of the scanning-NV microscope, {\it i.e.} the smallest distance for which distant point-like magnetic objects can be resolved. As discussed later in the paper, the resolving power is limited by the probe-to-sample distance. However, we stress that since the NV defect senses the field within an atomic-size detection volume, magnetic field distributions are recorded with nanoscale spatial resolution, even with a large probe-to-sample distance.\\
\indent In view of comparing the recorded magnetic image to micromagnetic predictions, the equilibrium magnetization distribution inside the thin disk of permalloy was computed using OOMMF software~\cite{oommf}. For this calculation, we used a saturation magnetization $M_s=8 \times 10^{5}$~A/m, an exchange constant $A=10^{-11}$~J/m and a magnetization cell size $2\times 2\times 2 \ {\rm nm}^{3}$, which is below the micromagnetic exchange length in order to describe the vortex core structure precisely. From the calculated equilibrium magnetization, the stray magnetic field was computed by summing the contribution of all magnetization cells, and then projected along the NV defect axis in order to reach a map of $B_{\rm NV}$~\cite{Rondin2013}. The only a priori unknown parameter of the simulation is the probe-to-sample distance $d$, since the position of the NV probe with respect to the apex of the AFM tip is not precisely controlled. This parameter was indirectly inferred by calculating the stray field at different distances $d$ and by choosing the one that gives the same maximum value for $|B_{\rm NV}|$ as in the experiment, {\it i.e.} $1.3$~mT in the conditions of Figure~\ref{Fig1}(e). This criterion relies on the strong distance dependence of the stray field above the vortex core. We found that a probe-to-sample distance $d=90$ nm leads to a simulated stray field distribution with a maximum value for $|B_{\rm NV}|$ of $1.3$~mT, as in the experiment, while for instance $d=80$ nm and $d=100$ nm would give about $1.7$~mT and $1.0$~mT, respectively~\cite{Sup}. Figures \ref{Fig1}(f) to (h) show the simulated images corresponding to the experimental images depicted in Figure~\ref{Fig1}(c) to (e) assuming $d=90$ nm. We note that the asymmetry observed in the simulations results from the imperfect alignment of the NV defect quantization axis with respect to the $z$-axis [Fig.~\ref{Fig1}(a)]. The calculated magnetic field distribution is in general agreement with the experimental data, with a stray field spread over $\approx 100$~nm. However, the magnetic image also reveals slight deviations from an ideal vortex structure~\cite{Yu2011}. We attribute these variations to imperfections of the magnetic structure, as discussed later in the article. \\

\begin{figure}[t]
\begin{center}
\includegraphics[width=0.49\textwidth]{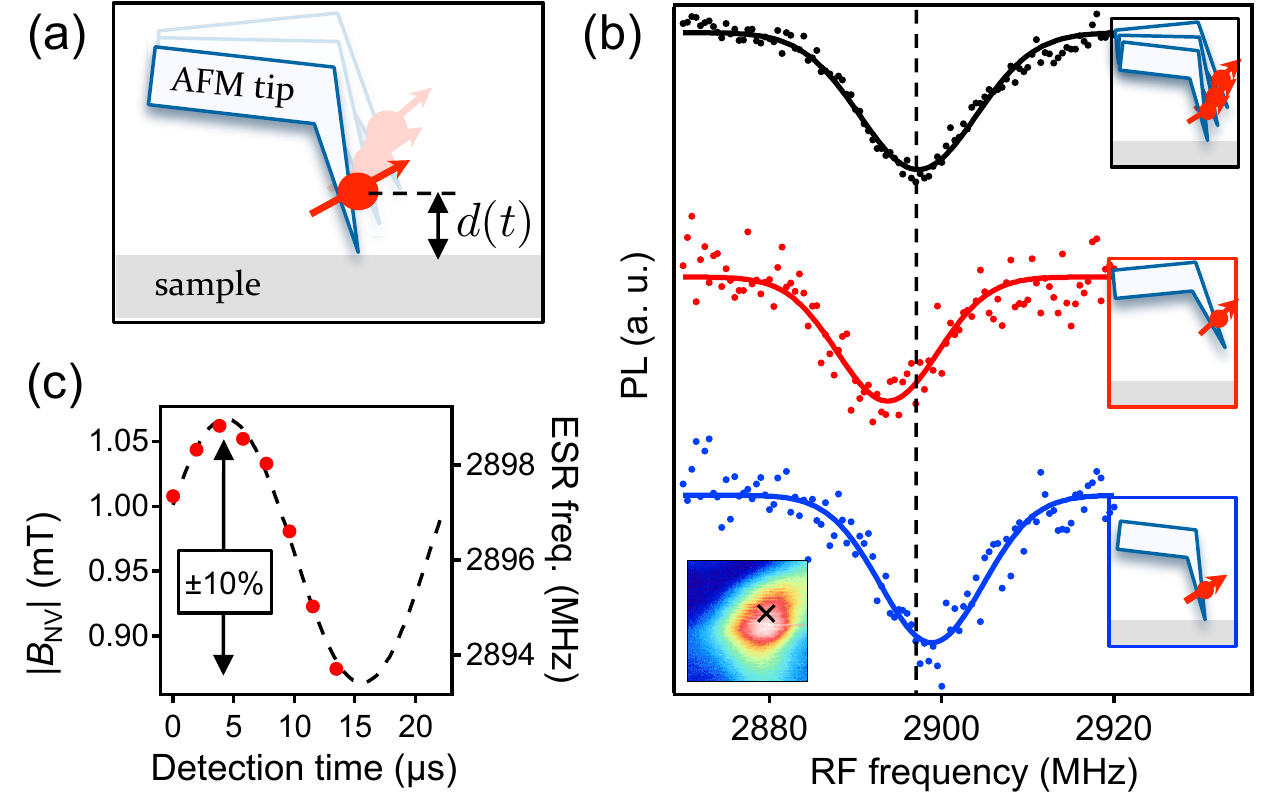}
\caption{(color online) (a) Schematic view of the tip oscillatory motion. (b) ESR spectra of the NV center, {\it i.e.} PL intensity versus RF frequency, recorded when the tip is placed above the vortex core (see cross in the inset). The spectrum depicted in the top panel is obtained by averaging over the tip oscillation. The two other spectra are obtained by synchronizing the acquisition with the tip position. The middle (resp. lower) panel corresponds to an ESR spectrum snapshot with the tip placed at its uppermost (resp. lowermost) position. Solid lines are gaussian fits to the data. (c) ESR frequency and corresponding magnetic field as a function of the detection time within the tip oscillation period. The dashed line is a guide-to-the-eye sine function at the frequency of the tip oscillation (45 kHz). The relative magnetic field variation is around $10\%$.}
\label{Fig2}
\end{center}
\end{figure}

\indent The probe-to-sample distance inferred from comparisons with micromagnetic simulations might seem surprisingly large given the 40-nm size of the diamond nanocrystal, measured by AFM prior grafting onto the tip. Two effects can explain this observation, as illustrated in Figure~\ref{Fig2}(a): (i) the diamond nanocrystal may actually be placed well above the end of the tip; (ii) the AFM tip, operated in tapping mode, oscillates vertically, resulting in a time-dependent distance $d(t)$ with a mean value $ \overline{d(t)}>0$. To quantify the latter effect, the acquisition of ESR spectra was synchronized with the tip oscillation~\cite{Hong2012}, whose frequency is about $45$~kHz. For these measurements, the tip was placed above the vortex core and the spin-dependent PL intensity of the NV defect was stroboscopically recorded within a detection window of duration $2 \ \mu$s, much shorter than the tip oscillation period, while sweeping the frequency of the RF field. The resulting ESR spectra are shown in Figure~\ref{Fig2}(b). The ESR frequency $\nu_r$ is shifted towards a higher value when the tip is at its lowermost position (bottom panel) compared with the uppermost position (middle panel), indicating a change of the local magnetic field $|B_{\rm NV}|$ within the tip oscillation. For comparison, the ESR spectrum shown in the top panel is obtained without synchronization, leading to averaging over the magnetic field modulation induced by the tip oscillation. In Figure~\ref{Fig2}(c), the magnetic field $|B_{\rm NV}|$ extracted from tip-synchronized ESR spectra is plotted as a function of the detection time within the tip oscillation period. A sinusoidal behavior is observed, which is the signature of the tip oscillation. In order to infer an estimate of the oscillation amplitude, we consider the vortex core as a point-like magnetic dipole with a stray field scaling as $d^{-3}$. This assumption is valid as long as the probe-to-sample distance remains significantly larger than the magnetic sample thickness ($t=50$~nm). The relative variations of $d$ and $|B_{\rm NV}|$ within the tip oscillation are then linked by $\delta B_{\rm NV}/B_{\rm NV}=3 \ \delta d/d$. The measured $10\%$ variation of the stray magnetic field  therefore translates into a $\approx 3\%$ oscillation of the probe-to-sample distance. Thus, the data shown in Figure~\ref{Fig1} corresponds to a mean probe-to-sample distance of $\approx 90$ nm with an oscillation amplitude $\approx 3$ nm. For magnetometry, the main consequences of this oscillation are (i) a broadened ESR line with reduced contrast which degrades the measurement sensitivity~\cite{Dreau2011}, and (ii) an increased mean tip-to-sample distance. However, in the present case these contributions are weak and the $90$-nm mean distance is attributed to imperfect positioning of the diamond nanocrystal at the end of the AFM tip. Note that the tip oscillation could also be used as a resource, either to extract three-dimensional information in a single scan~\cite{Tisler2013,Schell2013}, or to implement high-sensitivity AC magnetometry protocols~\cite{Hong2012}.\\
\indent The resolving power of the microscope could be significantly improved through precise control of the NV position at the end of the tip, {\it e.g.} by using diamond nanopillar probes~\cite{Maletinsky2012}. However, bringing the NV sensor very close to the sample surface would require to measure magnetic fields in the Tesla range for common ferromagnetic samples, such as the vortex structures investigated in this work. At such high fields, magnetometry based on measurements of Zeeman shifts of the NV defect electron spin sublevels is unpractical since the field is strong enough to induce electron spin mixing~\cite{Rondin2012}. Indeed, NV-based magnetometry is intrinsically limited to magnetic fields with an amplitude and an orientation such that the electron spin quantization axis remains fixed by the NV defect axis itself, corresponding to magnetic field amplitudes lower than few tenths of mT~\cite{Tetienne2012}. Consequently, the ultimate resolving power of scanning-NV magnetometry is a compromise between small probe-to-sample distance and sensing fields lower than few tenths of mT.\\

\begin{figure}[t]
\begin{center}
\includegraphics[width=0.48\textwidth]{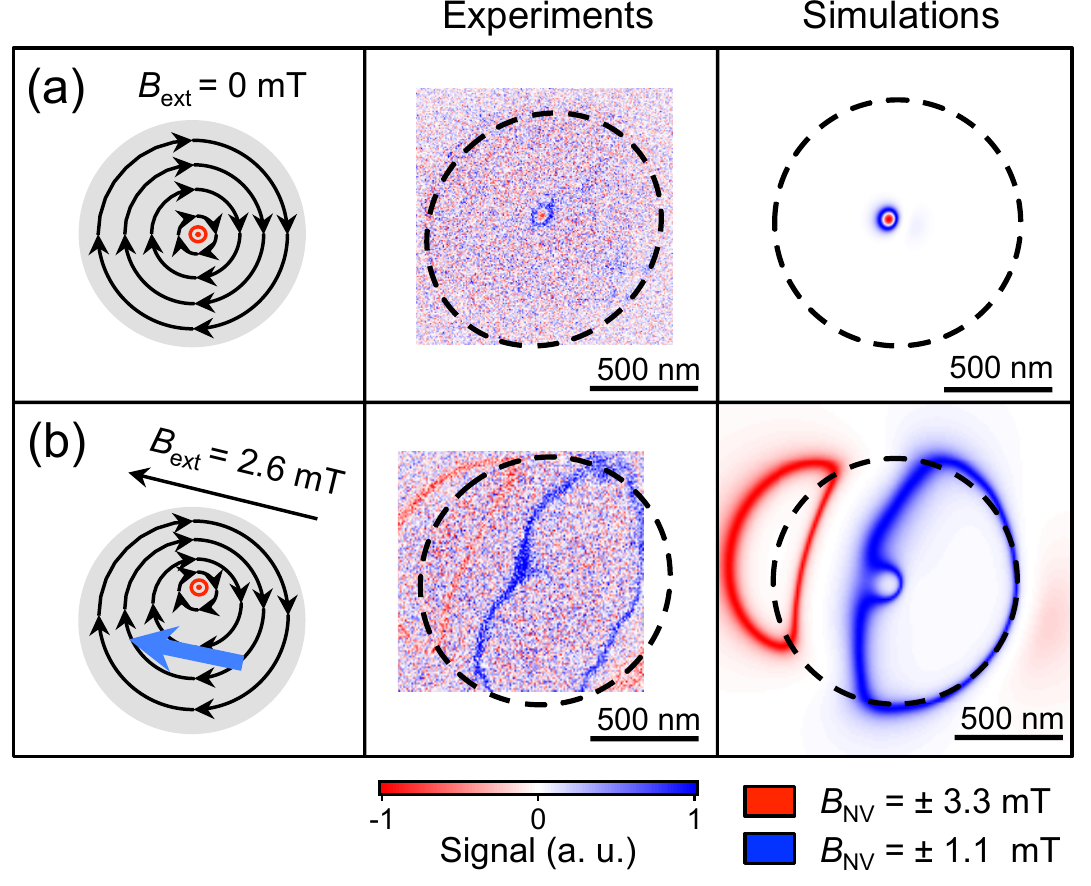}
\caption{(color online) (a) Measured and simulated dual-iso-B images in zero magnetic field and (b) with an in-plane bias field of $2.6$~mT. In the left panels, the black arrows indicate the curling magnetization of the vortex state and the red dot shows the out-of-plane magnetization of the vortex core, which is moved away from the center of the structure when a bias field is applied. The blue arrow depicts the net in-plane magnetization induced by vortex core displacement. The middle and right panels show the measured and simulated magnetic field images, respectively. The bias magnetic field is experimentally aligned along the in-plane projection of the NV center's quantization axis. Integration time per pixel: 60 ms for (a), 40 ms for (b).  }
\label{Fig3}
\end{center}
\end{figure}

\indent Next, we applied an in-plane bias magnetic field to the vortex structure and investigated its effect on the stray field. With a $2.6$~mT bias field, the vortex core is displaced away from the dot center, perpendicularly to the field direction, by $\approx 20$ nm according to micromagnetic simulations. Even though such a small shift is hard to observe directly, the vortex core displacement breaks the symmetry of the magnetization structure, which results in a net in-plane magnetization parallel to the bias field and delocalized over the whole disk. This in-plane magnetic moment produces a stray field that adds up to that of the vortex core. Figure~\ref{Fig3} shows dual-iso-B images recorded with and without the in-plane bias field. When a bias field is applied [Fig~\ref{Fig3}(b)], a two-lobe pattern spread over the whole disk is observed, as expected for an in-plane magnetic moment. This is in clear contrast with the zero-bias-field case, for which the stray field, which comes exclusively from the vortex core, is highly localized at the center of the disk [Fig~\ref{Fig3}(a)]. The corresponding simulated images reproduce well the dramatic change induced by the in-plane bias field. These measurements therefore reveal indirectly the nanoscale vortex core displacement induced by the bias field. 

\begin{figure}[h!]
\begin{center}
\includegraphics[width=0.48\textwidth]{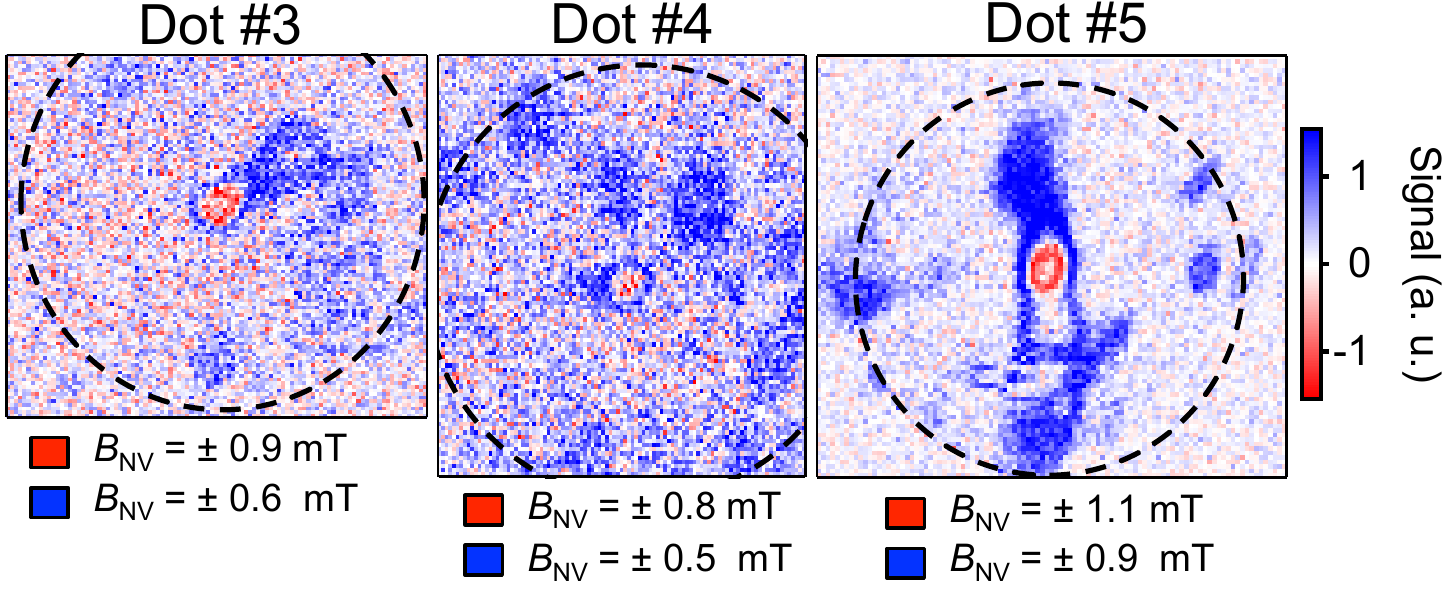}
\caption{(color online) Dual-iso-B images recorded above several nominally identical ferromagnetic dots with the same NV sensor. The dashed circles indicate the ferromagnetic dot boundaries, as extracted from AFM images. Integration time per pixel: 20 ms, 40 ms, 200 ms, for dots \#3, \#4, \#5, respectively.}
\label{Fig4}
\end{center}
\end{figure}

\indent Finally, several other ferromagnetic dots were imaged with the same NV sensor. Figure~\ref{Fig4} shows dual-iso-B images for three different dots. In each case, the vortex core is clearly distinguishable near the dot center, highlighted by the high-value iso-B contour (in red). On the other hand, the low-value contours ($\leq 0.9$ mT) reveal additional features in the stray field measured above the dots, which appear to be randomly distributed over the structure. Since those features differ from dot to dot, we attribute them to imperfections of the fabricated structures, {\it e.g.} edge and surface roughness. For instance, dot \#3 exhibits a spot of stray field next to the vortex core that reaches $\approx 0.6$ mT in value. This unexpected spot is likely due to some defect that either skews the vortex core or breaks the symmetry of the in-plane vortex distribution, producing additional stray field that interferes with the intrinsic vortex core stray field. This may explain the distortion observed in Figure~\ref{Fig1}(d),(e).

\indent In conclusion, we used a scanning NV-center microscope to quantitatively map the stray magnetic field above the vortex core of a ferromagnetic microdisk. This work demonstrates the effectiveness of NV-based magnetometry to obtain quantitative information on the stray field above magnetic nanostructures with high spatial and field resolution. In particular, weak perturbations with respect to the perfect vortex structure have been clearly detected, due to an applied field or imperfections of the magnetic samples. This sensitivity to weak magnetic fields, at the nanoscale, opens the door to the detection of exotic magnetic structures like skyrmions in ultrathin films~\cite{Heinze2011,Fert2013}.\\

\indent The authors acknowledge O. Klein for fruitful discussions. This work was supported by the Agence Nationale de la Recherche (ANR) through the project D{\sc iamag},  by C'Nano \^Ile-de-France (contracts M{\sc agda} and N{\sc anomag}) and by the European Union (project D{\sc iadems}).

\section*{Supplementary figure}

\begin{figure}[h!] 
\begin{center}
\includegraphics[width=7cm]{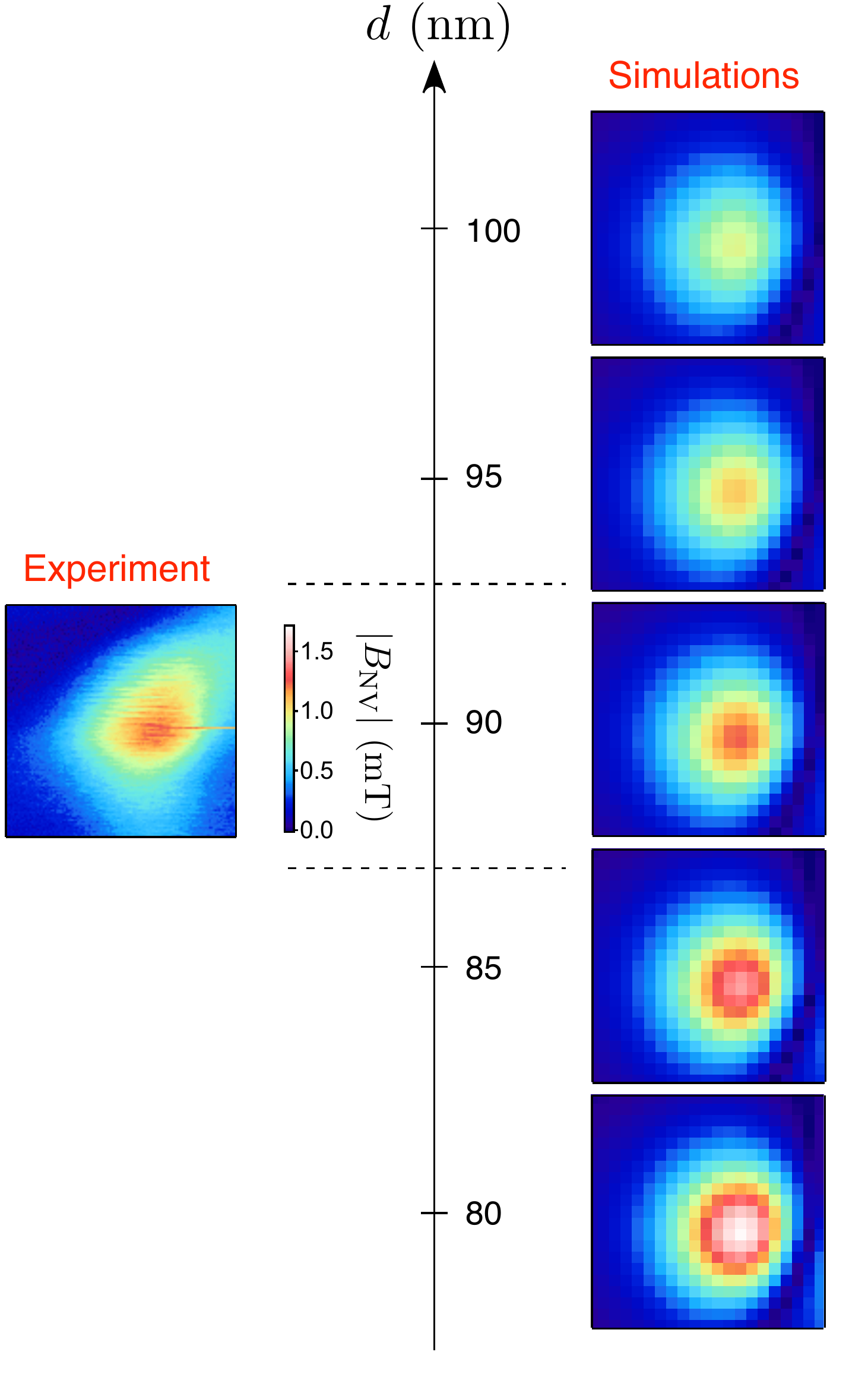}
\caption{Calculated stray magnetic field distribution above the vortex core for various probe-to-sample distances $d$. For $d=90$ nm, the simulated stray field distribution has a maximum value for $|B_{\rm NV}|$ of $1.3$~mT, as in the experiment, while for instance $d=80$ nm and $d=100$ nm give about $1.7$~mT and $1.0$~mT, respectively. }
\label{FigS1}
\end{center}
\end{figure}
\end{document}